\documentclass[pra,aps,twocolumn,showpacs,superscriptaddress]{revtex4-1}

\usepackage{epsfig}
\usepackage{mathrsfs}
\usepackage{amsmath}
\usepackage{float}
\usepackage{ulem}
\usepackage[usenames,dvipsnames]{color}

\begin{document}

\title{Counterfactual Logic Gates}

\author{Zheng-Hong Li}
\email{refirefox@shu.edu.cn}
\affiliation{
	Department of Physics, Shanghai University, Shanghai 200444, China}

\affiliation{
	Shanghai Key Laboratory for High Temperature Superconductors, Shanghai University, Shanghai 200444, China}

\author{Xiao-Fei Ji}
\affiliation{
	Department of Physics, Shanghai University, Shanghai 200444, China}

\author{Saeed Asiri}
\affiliation{CQOQI, KACST, P.O.Box 6086, Riyadh 11442, Saudi Arabia}
\affiliation{The National Center for Laser and Optoelectronics, KACST, Riyadh 11442, Saudi Arabia}

\author{Luojia Wang}
\affiliation{
State Key Laboratory of Advanced Optical Communication Systems and Networks, School of Physics and Astronomy, Shanghai Jiao Tong University, Shanghai 200240, China}

\author{M. \surname{Al-Amri}}
\affiliation{CQOQI, KACST, P.O.Box 6086, Riyadh 11442, Saudi Arabia}

\affiliation{The National Center for Laser and Optoelectronics, KACST, Riyadh 11442, Saudi Arabia}

\affiliation{Department of Physics, KKU, P.O. Box 9004, Abha 61413, Saudi Arabia}

\date{\today}

\begin{abstract}

We present how basic logic gates including NAND, NOR and XOR gates can be implemented counterfactually. The two inputs (Bob and Charlie) and the output (Alice) of the proposed counterfactual logic gate are not within the same station but rather separated in three different locations. We show that there is no need to pre-arrange entanglement for the gate, and more importantly, there is no real physical particles traveling among Alice, Bob and Charlie during the information processing. Bob and Charlie only need to independently control the blocking and unblocking of the transmission channels that connect them to Alice. In this way, they can completely determine the state of a real photon at Alice's end, thereby leading to implement a counterfactual logic gate. The functionality of a particular counterfactual logic gate is determined only by an appropriate design of Alice's optical device. Furthermore, by utilizing the proposed counterfactual logic gates, we demonstrate how to counterfactually prepare the Greenberger-Horne-Zeilinger state and W state with three remote quantum objects, which are in superposition states of blocking and unblocking the transmission channel.

\end{abstract}

\maketitle

\newcommand{\ds}{\displaystyle}
\newcommand{\dd}{\partial}
\newcommand{\be}{\begin{equation}}
\newcommand{\ee}{\end{equation}}
\newcommand{\beq}{\begin{eqnarray}}
\newcommand{\eeq}{\end{eqnarray}}
\newcommand{\dt}{\ds\frac{\dd}{\dd t}}
\newcommand{\dz}{\ds\frac{\dd}{\dd z}}
\newcommand{\D}{\ds\left(\frac{\dd}{\dd t} + c \frac{\dd}{\dd z}\right)}

\newcommand{\w}{\omega}
\newcommand{\W}{\Omega}
\newcommand{\g}{\gamma}
\newcommand{\G}{\Gamma}
\newcommand{\E}{\hat E}
\newcommand{\s}{\sigma}
\newcommand{\bra}{\langle}
\newcommand{\ket}{\rangle}

\section{Introduction}

Quantum measurement plays a very important role in quantum mechanics, and brings many fascinating results,  such as interaction free measurement \cite{inf1,inf2} and quantum Zeno effect \cite{QZ1,QZ2,inf3}, which leads to the fact that frequent measurements of a quantum system inhibits its unitary evolution. Utilizing this fact, the direct counterfactual quantum communication protocol (SLAZ) was proposed theoretically \cite{ours2}, and successfully verified experimentally \cite{exp01,exp02}. Such protocol shows that information can be transmitted without any physical particles traveling between two communicating parties, Alice and Bob. The only action Bob needs to take is to control the blocking or unblocking of the transmission channel that connects him to Alice. By doing that, he is able to manipulate a single photon that is confined to Alice's station.

Potential applications of the direct counterfactual quantum communication protocol include detection of vulnerable samples \cite{exp01}, quantum secure communication \cite{CounterCry1}, quantum eavesdropping \cite{ours1,2020}, Bell-state analysis \cite{CB}, quantum cloning \cite{qcloning}, entanglement distribution \cite{entanglement1,CE,otherC2,entanglement2} and much more \cite{otherC1,CNOT,tele0,ours4,ourpra}.  It is worth noting that in the SLAZ protocol, the control of the transmission channel is implemented by a classical object to either block or unblock the transmission channel using for example a switchable detector. Instead of the classical object, one can have a quantum object  \cite{entanglement1,qcloning,CE,otherC2,entanglement2,CB,otherC1,CNOT,tele0,ours4,ourpra}, which can be in a superposition of blocking or unblocking the transmission channel. Such quantum object can be a Rydberg atom \cite{tele0,Atom1,Atom2}, a single-side cavity with a three-level atom \cite{ours4,qn27,qn28,inputout} and so on. It is shown that the quantum object can be entangled with Alice's photon without any photon traveling between them \cite{entanglement1,tele0,otherC2,entanglement2,CE}. Based on the achieved counterfactual entanglement, 2-qubit quantum gates such as CNOT gate \cite{CNOT,CB}, controlled-phase gate \cite{otherC1} and quantum swap gate \cite{ourpra} have been investigated and accomplished counterfactually.

  Up to now, most of related works focus on only two remote parties. We notice that when considering network applications, research on multiparty counterfactual quantum control is necessary. Therefore, in this paper, we go further and study the situation involving three or more parties. More specifically, only one party holds an optical device with a real photon inside it, while other parties independently control the transmission channels utilizing either classical or quantum objects. In case that classical objects are used, we show that the photon can be manipulated without any need for pre-arranged entanglement or any real physical particles traveling among those parties. Assuming that the status of the classical object represents input while the final state of the photon represents the output. The exclusive design of the optical device can lead to counterfactually implementation of NAND, NOR and XOR gates. This is the main contribution of this work. Moreover, based on those counterfactual logic gates, we show that when classical objects are replaced by quantum objects, it is possible to counterfactually entangle three remote quantum objects with the assistance of the fourth parity's photon, and hence realize Greenberger-Horne-Zeilinger (GHZ) state and W state \cite{GHZ1,GHZ3,GHZ2}. Our results will facilitate the study of 3-qubit counterfactual quantum gates including Toffoli and Fredkin gates, along with research on counterfactual multi-party entanglement.

The structure of this paper is as follows. In Section II,  we introduce the basic model and calculation method of the present work. In Section III, we present the counterfactual NAND gate. Since its design is very close to the SLAZ protocol, this section can be regarded as a brief review of the SLAZ protocol. In Section IV, we introduce the counterfactual logic gate unit (CGU), which is an essential component of the proposed NOR gate and XOR gate. In Section V, we present the scheme of the counterfactual NOR gate along with numerical simulation. In Section VI, we present the counterfactual XOR gate and its numerical simulation results. In Section VII, we show how to use counterfactual logic gates to entangle three quantum objects (V-type three-level atom), thus realizing GHZ state and W state. In Section VIII, we give a short discussion about the experiment. In Section IX, we present our concluding remarks.

\section{Basic model and calculation method}

\subsection{Tripartite model}

\begin{figure}[htbp]
	\centering\includegraphics[width=1.0\columnwidth]{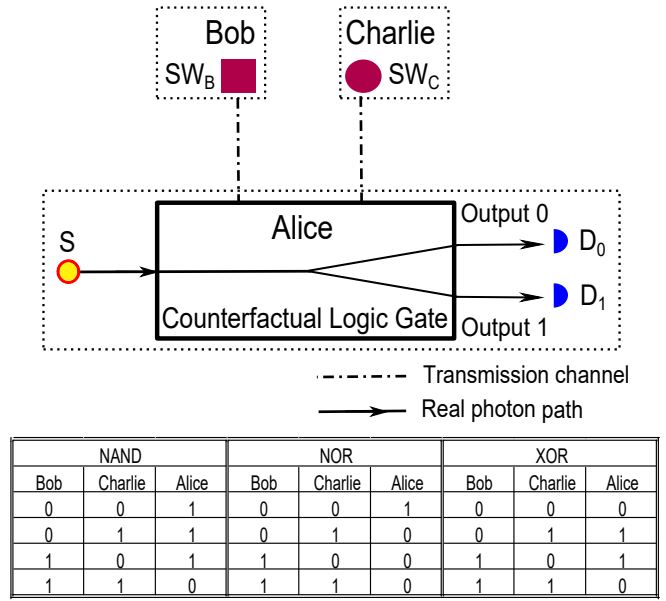}
	\caption
	{ \label{fig:fig1}
		The basic model of the counterfactual logic gates and their corresponding truth tables. Alice, Bob and Charlie are three remote parties. At Bob's (Charlie) end, he blocks or unblocks the transmission channel that connects him to Alice by means of switchable detector $SW_{B(C)}$. Blocking represents input 0, while unblocking represents input 1. At Alice's end, $S$ stands for single photon source, and $D$ stands for detector. Inside Alice's box, there is an optical device with specific design to suit the chosen counterfactual logic gate. The details are presented in Figs. \ref{fig:fig3}, \ref{fig:fig4} and \ref{fig:fig7}. Initially, a single photon is sent into the device but its path (solid lines) is different according to Bob and Charlie's inputs. If the photon is routed to detector $D_0$, that represents output 0. But if it is routed to $D_1$, that represents output 1.}
\end{figure}

In Fig.\ref{fig:fig1}, Alice, Bob and Charlie are three remote parties. At Alice's end, we have a box that contains optical design for counterfactual logic gate. It consists of beam-splitters ($BS$), normal mirrors ($MR$) and detectors ($D$) (Not shown in the figure). Initially, a single photon is generated by a single photon source ($S$) and enters Alice's device with a fixed initial state. After that, its real photon path (the solid lines) is determined by the design of Alice's device and the actions of Bob and Charlie. Regarding Bob and Charlie, they have separate transmission channels (the dotted dashed lines) connecting them to Alice. The actions, they take, are to block or unblock their own transmission channels. In experiment, these actions can be achieved with switchable detector ($SW$), which is a classical object. If the $SW$ is turned on, it becomes a single photon detector. If it is off, it becomes a mirror and returns photons entering the transmission channel to Alice's device. In that sense, the transmission channel is unblocked. Here, we define these actions as inputs to the counterfactual logic gate. More specifically, blocking (unblocking) the transmission channel represents logic 0 (1). According to Bob and Charlie's actions, Alice's photon is routed to different directions. We define the final photon states as outputs to the counterfactual logic gate. More specifically, if the photon is routed to output 0(1), that represents logic 0(1). 

Based on the above definitions, with specific designs of Alice's device, we can implement the counterfactual NAND, NOR and XOR gates whose truth tables are shown in Fig.\ref{fig:fig1}. The details will be discussed later. Here, we only emphasize three issues:

(1) In the detailed designs (see Fig.\ref{fig:fig3}, Fig.\ref{fig:fig4} and Fig.\ref{fig:fig7}), except for $SW_B$s (controlled by Bob) and $SW_C$s (controlled by Charlie), all other components belong to Alice.

(2) Due to the quantum Zeno effect, the proposed counterfactual logic gates require Bob and Charlie to ``manipulate'' Alice's photon many times to obtain the expected output. Therefore, there are many $SW_B$s and $SW_C$s in our designs. However, the action of all $SW_B$s are consistent, and so are  the $SW_C$s.

(3) Since initial Alice's photon state is fixed, the proposed logic gates are not quantum gates. But this does not mean that we cannot use these logic gates to achieve some quantum phenomena such as 3-qubit counterfactual quantum entanglement.

\subsection{Beam-splitter}	

When designing Alice's device, a key optical component is the beam-splitter, which is denoted by $BS^{(K)}_k$. Here, the superscript represent the type of the $BS$, while the subscript $k$ represents that $BS^{(K)}_k$ is the $k$-th $K$-type $BS$. Suppose now that the state $\left|L_K\right\rangle$ represents a photon on the left side of the $BS^{(K)}$, while the state $\left|R_K\right\rangle$ for a photon on the right side of the $BS^{(K)}$. The function of $BS^{(K)}$ can be represented as

\begin{equation}
\begin{split}
\label{e1}
&|L_K\rangle \to \cos \theta_K |L_K\rangle +\sin \theta_K |R_K\rangle   \\
&|R_K\rangle \to \cos \theta_K |R_K\rangle -\sin \theta_K |L_K\rangle   \\
\end{split}
\end{equation}
Here, $\left|\cos \theta_K \right|^2$ is the reflectivity of the $BS^{(K)}$, while $\left|\sin \theta_K \right|^2$ is the transmissivity. Therefore, the superscript $K$ refers to the transmissivity of the $BS$.

In this work, there are totally 5 types of $BS$s (see Figs.\ref{fig:fig3}, \ref{fig:fig4} and \ref{fig:fig7}) which have different transmissivity. We use superscripts $2,M,N,A1,A2$ to distinguish them. For $BS^{(A1)}$ and $BS^{(A2)}$, their transmissivity are $\sin\theta_{A1}=\cos^{2N}(\pi/2N)$ and $\sin\theta_{A2}=\cos^{6N}(\pi/2N)$, respectively. These two types of $BS$s are used for attenuators \cite{Pnoise}. $BS^{(A1)}$ and $D_{A1}$ form Attenuator 1 (see Fig. \ref{fig:fig3}) while $BS^{(A2)}$ and $D_{A2}$ form Attenuator 2 (see Fig. \ref{fig:fig4}). After passing through Attenuator 1(2), a photon has a $\sin^2\theta_{A1(2)}$ probability of remaining in Alice's device. In section V, we will show these attenuators are essential to the counterfactual NOR gate. As for $BS^{(2)}$, $BS^{(M)}$ and $BS^{(N)}$, their transmissivity is $\theta_{K}=\pi/2K$ ($K=2,M,N$, where $M$ and $N$ are integers). Apparently, $BS^{(2)}$ has equal transmissivity and reflectivity. As for $BS^{(M)}$ and $BS^{(N)}$, we require their transmissivity to be small. They are used to build up the interferometer chain, which will be discussed below.

\subsection{Chained interferometers}	

\begin{figure}[htbp]
	\centering\includegraphics[width=0.95\columnwidth]{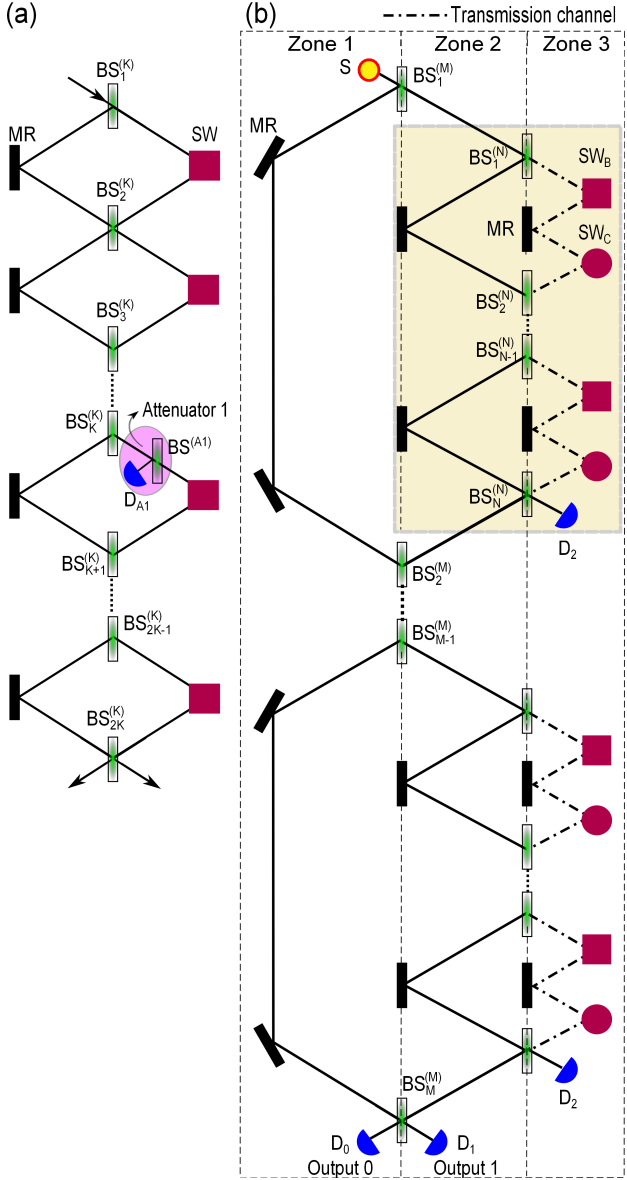}
	\caption
	{ \label{fig:fig3}
		(a) The chained Mach-Zehnder interferometers composed of $2K$ beam-splitter $BS^{(K)}$. $MR$ stands for normal mirror. The right arms of the chain are controlled by $SW$s. If the $SW$ is turned on, it becomes a detector. If it is off, it becomes a mirror so that the interference in the chain is not interrupted. In addition, the attenuator is composed of $BS^{(A1)}$ and $D_{A1}$. (b) The scheme of the counterfactual NAND gate. It is similar to the SLAZ protocol, except that the transmission channel is shared and controlled by two parties Bob ($SW_B$) and Charlie ($SW_C$).}
\end{figure}

The optical structure of the interferometer chain is essential for all proposed counterfactual logic gate designs. In general, the chain is made up of many Mach-Zehnder interferometers that are connected in series. With appropriate settings, we can use that to manipulate a photon such as its path or phase. For the convenience of discussion, we show an example in Fig.\ref{fig:fig3}(a). This chain has $2K$ number of $BS^{(K)}$s with $K=M,N$. Accordingly, there are $2K-1$ Mach-Zehnder interferometers. The right arms of these interferometers are controlled by $SW$s, whose actions are consistent. If $SW$s are turned on (off), these arms are blocked (unblocked), which we call it the blocking (unblocking) case. In addition, in the figure, the Attenuator 1 is on the right side of $BS^{(K)}_K$, which, however, is optional. In fact, by adjusting the total number of $BS^{(K)}$, adding the attenuator, and setting $SW$s, we can achieve four different types of photon control. In the following, we explain that based on Fig.\ref{fig:fig3}(a) and then we summarize our conclusion into four working modes.

Assuming that initially a single photon is sent to the chain from the left side. The corresponding initial photon state is $|L_K\rangle$.  

First, we discuss the unblocking case. If the attenuator is not considered, it is easy to see that after $k$-th $BS^{(K)}$, the photon state becomes
\begin{equation}
\begin{split}
\label{d1}
\cos(\pi k/2K)|L_K\rangle +\sin(\pi k/2K)|R_K\rangle.
\end{split}
\end{equation} 
Apparently, when $k=K$, the photon state is $|R_K\rangle$, and the photon appears on the right side. When $k=2K$, the photon state is $-|L_K\rangle$, and the photon remains on the left side but with a $\pi$ phase shift. 

If the attenuator is considered, the evolution of the photon state can be calculated as follows. We notice that after $BS^{(K)}_K$, the photon state is $|R_K\rangle$. Regarding Attenuator 1, its contribution can be described as $|R_K\rangle \to \cos^{2N}(\pi/2N)|R_K\rangle$ when $D_{A1}$ does not find the photon. Here (and in the rest of this paper), we do not do renormalization to facilitate the calculation of probability. It is easy to see that the probability of the photon remaining in the chain is $\cos^{4N}(\pi/2N)$. Then, after another $K$ $BS^{(K)}$s, the final photon state becomes $-\cos^{2N}(\pi/2N)|L_K\rangle$ according to Eq.\eqref{e1}. 

Second, we discuss the blocking case. After $BS^{(K)}_k$, the photon state is ${\cos}^k(\pi/2K)\left|L_K\right\rangle+{\cos}^{k-1}(\pi/2K)\sin(\pi/2K)\left|R_K\right\rangle$, which is correct under the condition that no photon is found by any $SW$s. When $K$ is large, the probability of the photon being found on the left side of the chain is ${\cos}^{2k}(\pi/2K) \approx 1-\pi^2k/4K^2$, which tends to one when $K^2 \gg 2 k$. In this scenario, it is also easy to see that the attenuator has no contribution to the result.

To simplify the discussion in the following sections, we summarize four working modes of the interferometer chain with $K=M,N$. 

{\bf{Working mode 1: the blocking case.}}

The photon state evolution after $k$ $BS^{(K)}$s can be approximately written as $|L_K\rangle \to {\cos}^k(\pi/2K)\left|L_K\right\rangle$.

{\bf{Working mode 2: the chain contains $K$ $BS^{(K)}$s; the unblocking case.}}

The photon state evolution can be represented as $|L_K\rangle \to |R_K\rangle$. The photon is routed from the left side to the right side of the chain.

{\bf{Working mode 3: the chain contains $2M$ $BS^{(M)}$s without the attenuator; the unblocking case.}}

The photon state evolution can be represented as $|L_M\rangle \to -|L_M\rangle$. The photon remains on the left side of the chain, but with a $\pi$ phase shift.

{\bf{Working mode 4: the chain contains $2N$ $BS^{(N)}$s with Attenuator 1 in the middle; the unblocking case.}}

 Attenuator 1 is located on the right side of the $N$-th interferometer. The photon state evolution can be represented as $|L_N\rangle \to -\cos^{2N}(\pi/2N)|L_N\rangle$.

\section{Counterfactual NAND gate}

In this section, we introduce the counterfactual NAND gate, whose truth table is shown in Fig.\ref{fig:fig1}. Since its basic idea is similar to the SLAZ protocol, here we only present a brief discussion. For more analyses and details, including error analysis, one can consult  previous studies \cite{ours2}.

\subsection{Scheme}

We present the scheme of the counterfactual NAND gate as in Fig.\ref{fig:fig3}(b). It is very close to the SLAZ protocol and also has the structure of double chained Mach-Zehnder interferometers. The outer chain contains $M$ $BS^{(M)}$s. They constitute $M-1$ outer interferometers. On the right arm of each outer interferometer, there are $N$ $BS^{(N)}$s, which is called the inner chain and contains $N-1$ inner interferometers. The only difference between the present scheme and the SLAZ protocol is about how to control the transmission channel, i.e., the right arms of the inner interferometers. As shown in the figure, each of these arms is controlled by both Bob and Charlie. Any one of them selects to block, the arm is blocked. In other words, we can regard that Bob and Charlie share and control the same transmission channel, but in the SLAZ protocol, the channel is controlled by Bob only. 

For the convenience of discussion, we set three zones as in Fig. \ref{fig:fig3}(b) and assume that the photon state $|100\rangle$ represents a single photon in Zone 1 (the left side of $BS^{(M)}$s), $|010\rangle$ represents a single photon in Zone 2 (the area between $BS^{(M)}$s and $BS^{(N)}$s), and $|001\rangle$ represents a single photon in Zone 3 (the right side of $BS^{(N)}$s). The initial Alice's photon state is $|100\rangle$. Based on the different status of the transmission channel, the photon evolves differently. 

First, we consider the case that the transmission channel is unblocked. In such case, any photon entering the inner chains must be routed to $D_2$ (working mode 2), and it never returns to the outer chain. As a result, the outer chain is in working mode 1, and the final photon state approximately is $\cos^{M}(\pi/2M)|100\rangle$. The probability of $D_0$ clicking tends to one when $M\rightarrow\infty$. 

Second, we consider the case when the transmission channel is blocked. In such case, any photons entering the transmission channel must be absorbed by $SW$s. As a result, the inner chains are in working mode 1. In case $N$ is sufficiently large, we can regard that the interference in the outer chain is not interrupted. In other words, the outer chain is in working mode 2, which leads $D_1$ to click. As for the probability of $D_1$ clicking, we can approximately calculate it as follows. Still assume that $N$ is sufficiently large and the outer chain is in working mode 2. Due to Eq.\eqref{d1}, the probability of Alice's photon entering the inner chain in the $m$-th outer interferometer is ${\sin}^2(m\pi/2M)$. We notice that in each inner chain, the probability of a photon being detected in the transmission channel is $1-\cos^{2N}\left(\pi/2N\right)$. This leads to the total probability of Alice's photon being found in the transmission channel as \cite{ouroe}  
\begin{equation}
\begin{split}
\label{e3}
\sum_{m=1}^{M}{{\sin}^2\frac{m\pi}{2M}}\left(1-{\cos}^{2N}\frac{\pi}{2N}\right)\approx\frac{M\pi^2}{8N}.
\end{split}
\end{equation}
Then, the probability of $D_1$ clicking approximately is $1-M\pi^2/8N$, which tends to 1 when $N \gg M$. 

In summary, when the transmission channel is unblocked (blocked), $D_{0(1)}$ clicks. In case that only Bob controls the transmission channel (the SLAZ protocol), according to the definition in Section II, a NOT gate is implemented. In case that Bob and Charlie control the transmission channel, the channel is blocked no matter who selects to block (input 0). In such cases, $D_{1}$ clicks (output 1), while only when both Bob and Charlie unblock (input 1) the transmission channel, $D_{0}$ clicks (output 0). Consequently, the truth table of the NAND gate is achieved.

\subsection{Analysis of counterfactuals}

Here, we explain why these gates are counterfactual. First of all, since the single photon is used, as long as any $SW$ or $D_2$ finds Alice's photon, $D_0$ and $D_1$ never click. Based on that fact, we discuss the blocking and unblocking cases, respectively. When the transmission channel is blocked, any photons appear in the transmission channel must be detected by $SW$s. If $D_1$ clicks, it means the photon never enters the transmission channel. When the transmission channel is unblocked, we have shown that any photon enters the inner chains must be routed to $D_2$. Then, we can treat each inner chain and the corresponding $D_2$ (see the yellow rectangle in Fig. \ref{fig:fig3}(b)) as a combined detector. It is easy to see that when $D_0$ clicks, the photon never triggers the combined detectors. In other words, it never enters the inner chains. We emphasize that the transmission channel is in the right arms of the inner chains. Thus, the photon never enters the transmission channel. As a result, the presented NOT gate and NAND gate are counterfactual.

\subsection{M-type counterfactual NAND gate}

At the end of this section, we should point out that if there are more than two parties controlling the transmission channel (say, Bob, Charlie and David), the same result can be obtained, i.e., only when all of them have input 1, Alice obtains the output 0. Otherwise, Alice has the output 1 since the transmission channel must be blocked, which is the case no matter who chooses to block the channel. We call such gate as the M-type counterfactual NAND gate.

	\section{Counterfactual logic gate unit (CGU)}
	
\begin{figure}[htbp]
	\centering\includegraphics[width=0.8\columnwidth]{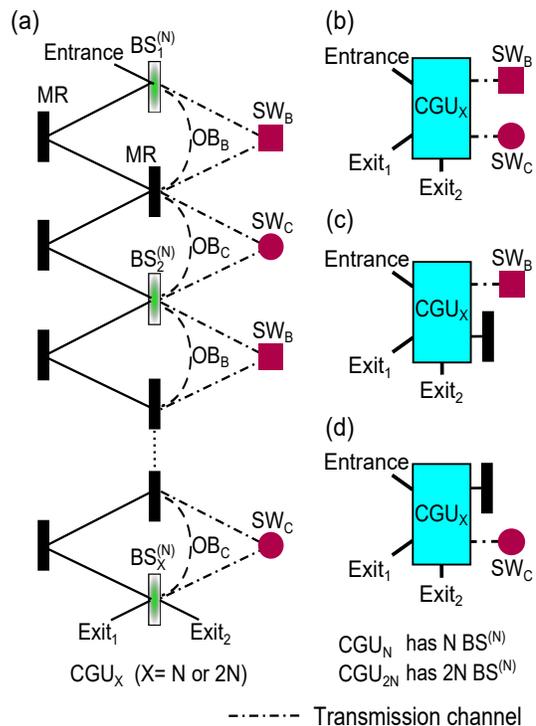}
	\caption
	{ \label{fig:fig2}
		(a) The details of the counterfactual logic gate unit (CGU$_\text{X}$), where the subscript $X$ represents there are $X$ number of $BS^{(N)}$s in the unit. Consequently, there are two types of this unit, CGU$_\text{N}$ and CGU$_\text{2N}$. In addition, $OB$ stands for optical bridge. If an $OB$ is activated, it routes the photon to skip the $SW$ on the right side. As a result, Alice can decide who controls the CGU$_\text{X}$ simply by utilizing $OB$s. (b) The simplified diagram of CGU$_\text{X}$ representing the case when both Bob and Charlie control the CGU$_\text{X}$. Any of them selects to block, the CGU$_\text{X}$ is in the blocking case. (c) The simplified diagram representing that only Bob controls the CGU$_\text{X}$. (d) The simplified diagram representing that only Charlie controls the CGU$_\text{X}$.}
\end{figure}
So far we present the counterfactual NAND gate, and when it comes to the counterfactual NOR and XOR gates, Alice's device needs to be redesigned. For this reason and to make the following discussion accessible, we introduce what we called counterfactual logic gate unit (CGU) in this section. The detailed scheme is presented in Fig.\ref{fig:fig2}(a), which has one photon entrance and two exits. The main structure is a chain of Mach-Zehnder interferometers formed by $BS^{(N)}$. Depending on the total number of $BS^{(N)}$, there are two types of CGUs. 

The first type is CGU$_\text{N}$, which contains $N$ $BS^{(N)}$. According to Section II, if its right arms are blocked, it is in working mode 1. If not, it is in working mode 2. 

The second type is CGU$_\text{2N}$. It consists of $2N$ $BS^{(N)}$ and an Attenuator 1 (not shown in Fig.\ref{fig:fig2}(a)). The location of the Attenuator 1 is on the right side of the chain and between $BS^{(N)}_N$ and $BS^{(N)}_{N+1}$. Therefore, CGU$_\text{2N}$ is in working mode 4 if its right arms are unblocked, while it is in working mode 1 if the arms are blocked. 

As for blocking or unblocking, these actions are performed by $SW_B$ and $SW_C$. We emphasize that $SW_B$ is controlled by Bob, and $SW_C$ by Charlie, but all other parts of CGU$_\text{X}$ ($X=N,2N$) belong to Alice. In addition, as shown in the figure, the transmission channel is in the right arms of the chained interferometers. If Alice does nothing, the transmission channel will be shared and controlled by both Bob and Charlie. However, at Alice's end, she can use optical bridges ($OB$) to bypass either Bob or Charlie's $SW$ so that the CGU$_\text{X}$ is controlled by either Charlie or Bob alone. As illustrated in Fig.\ref{fig:fig2}(b)(c)(d), we use simplified diagrams to represent the following three situations. Fig.\ref{fig:fig2}(b) represents the case when both Bob and Charlie control the CGU$_\text{X}$. Any one of them selects to block, the right arms of the CGU$_\text{X}$ are blocked. (c) represents the case that only Bob controls the CGU$_\text{X}$, while (d) represents that only Charlie controls the CGU$_\text{X}$.

\section{Counterfactual NOR gate}

In this section, we introduce the counterfactual NOR gate, whose truth table is shown in Fig.\ref{fig:fig1}. 

\subsection{Scheme}

\begin{figure}[htbp]
	\centering\includegraphics[width=0.8\columnwidth]{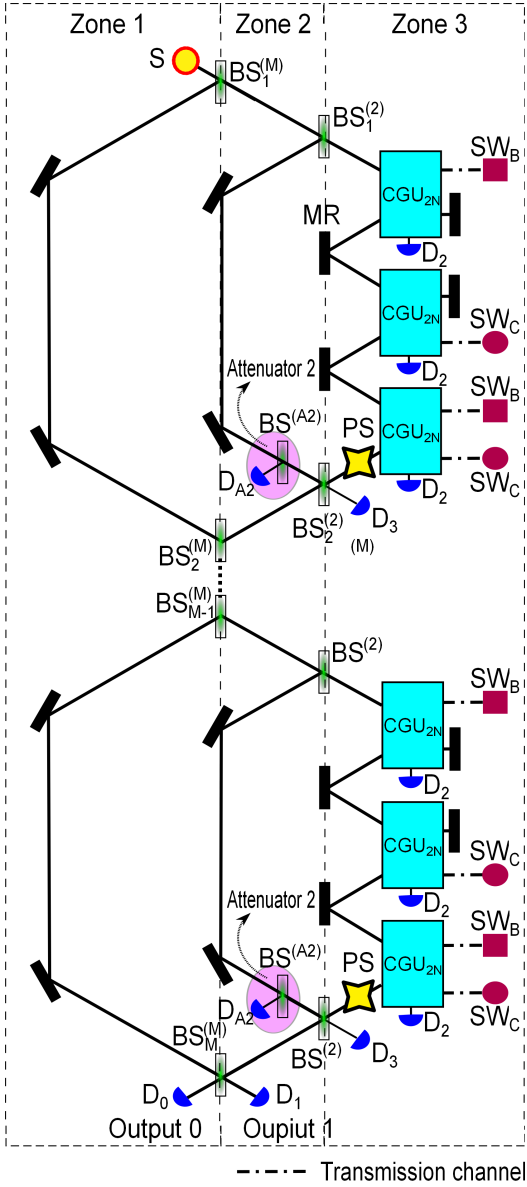}
	\caption
	{ \label{fig:fig4}
		The scheme of the counterfactual NOR gate. $PS$ stands for phase shifter. $BS^{(A2)}$ and $D_{A2}$ form Attenuator 2. As for the details of CGU$_\text{2N}$, one can consult  Fig.\ref{fig:fig2}. }
\end{figure}

In Fig.\ref{fig:fig4}, we present the detailed scheme, which has the structure of triple chained interferometers. The biggest chain consists of $M$ $BS^{(M)}$s, which we still call the outer chain. In the right arm of each outer interferometer, there is a Mach-Zehnder interferometer composed of two $BS^{(2)}$s. We call this interferometer ``the middle interferometer''. In the left arm of each middle interferometer, there is an Attenuator 2. In the right arm of each middle interferometer, there are three CGU$_\text{2N}$s in series, which are collectively called the inner chain. Each CGU$_\text{2N}$ has a detector ($D_2$) for measuring photon at exit 2, while at exit 1 of the third CGU$_\text{2N}$, there is a phase shifter ($PS$), which adds a $\pi$ phase shift to any photons passing through it. 

Now, we explain how the counterfactual NOR gate works. For clarity, we divide Fig.\ref{fig:fig4} into three zones. If we have a single photon in Zone 1, i.e., the left side of $BS^{(M)}$s, the corresponding photon state is $\left|100\right\rangle$. If a photon is in Zone 2, i.e., the area between $BS^{(M)}$s and $BS^{(2)}$s, the corresponding photon state is $\left|010\right\rangle$, while in Zone 3, i.e., the right side of $BS^{(2)}$s, the corresponding photon state is $|001\rangle$. 

Initially, a single photon in state $\left|100\right\rangle$ is sent into the counterfactual NOR gate. After passing through $BS^{\left(M\right)}_1$ and $BS^{\left(2\right)}_1$, the photon state becomes $\cos(\pi/2M)|100\rangle+\sin(\pi/2M)(|010\rangle+|001\rangle)/\sqrt{2}$. For the photon in state $|010\rangle$, it passes through Attenuator 2, which leads to $\left|010\right\rangle\rightarrow\cos^{6N}\left(\pi/2N\right)|010\rangle$. As for the photon in state $|001\rangle$, it passes through all the three CGU$_\text{2N}$s one by one. According to Bob and Charlie's different inputs, Alice's photon evolves differently.

{\bf{Case 1: Both Bob and Charlie have input 0.}}

In this case, all CGU$_\text{2N}$s are in working mode 1. The evolution of the photon state in an inner chain can be described as $\left|001\right\rangle\rightarrow{-\cos}^{6N}\left(\pi/2N\right)|001\rangle$, where the minus sign comes from $PS$. Then, the photon state before $BS^{\left(2\right)}_2$ is $\cos(\pi/2M)|100\rangle + \sin(\pi/2M)\cos^{6N}(\pi/2N)(|010\rangle-|001\rangle)/\sqrt{2}$. Here, we can see that due to Attenuator 2, the two arms of the middle interferometer are balanced. After passing through $BS^{\left(2\right)}_2$, the photon state becomes $\cos(\pi/2M) \left|100\right\rangle + \sin(\pi/2M) \cos^{6N} (\pi/2N) \left|010\right\rangle$. When $N$ is sufficiently large, the outer chain can be considered as being in working mode 2. Similar to the calculation of Eq.\eqref{e3}, we can estimate that the probability of $D_1$ clicking is
\begin{equation}
\begin{split}
\label{e5}
P_{00D1}=1-\sum_{m=1}^{M}{{\sin}^2\frac{m\pi}{2M}\left(1-\cos^{12N}\frac{\pi}{2N}\right)}\approx1-\frac{3\pi^2M}{4N}
\end{split}
\end{equation}
When $N\gg M$, Alice's photon triggers $D_1$ with almost 100\% probability. The corresponding output is 1.

{\bf{Case 2: Both Bob and Charlie have input 1.}}

In this case, all CGU$_\text{2N}$s are in working mode 4. The photon state evolution in an inner chain can be described as $\left|001\right\rangle\rightarrow\cos^{6N}\left(\pi/2N\right)|001\rangle$. Accordingly, the photon state before $BS^{\left(2\right)}_2$ is $\cos(\pi/2M)\left|100\right\rangle+\sin(\pi/2M)\cos^{6N}(\pi/2N)(|010\rangle+|001\rangle)/\sqrt2$. Here, the two arms of the middle interferometer are still balanced due to the use of the attenuators in CGU$_\text{2N}$s. However, comparing with case 1, a $\pi$ phase difference occurs. Then, after the photon passing through $BS^{\left(2\right)}_2$, the photon state becomes $\cos(\pi/2M)\left|100\right\rangle+\sin(\pi/2M)\cos^{6N}(\pi/2N)\left|001\right\rangle$. Here, the photon in state $|001\rangle$ is measured by $D_3$. Consequently, as for the outer chain, it is in working mode 1. After $M$ $BS^{(M)}$s, the probability of $D_0$ clicking is
\begin{equation}
\begin{split}
\label{e7}
P_{11D0}=\cos^{2M}\frac{\pi}{2M}\approx1-\frac{\pi^2}{4M}
\end{split}
\end{equation} 
When $M\rightarrow\infty$, Alice's photon triggers $D_0$ with almost 100\% probability. The corresponding output is 0.

{\bf{Case 3: Bob has input 0 and Charlie has input 1.}}

In this case, the first and third CGU$_\text{2N}$s in the inner chain are in working mode 1, while the second CGU$_\text{2N}$ is in working mode 4. Taking $PS$ into account, the evolution of a photon state in an inner chain can be described as $\left|001\right\rangle\rightarrow\cos^{6N}\left(\pi/2N\right)|001\rangle$. Similar to case 2, the probability of Alice's $D_0$ clicking (output 0) is $P_{01D0}=\cos^{2M}(\pi/2M)\approx1-\pi^2/4M$. The probability tends to one when $M\rightarrow\infty$.

{\bf{Case 4: Bob has input 1 and Charlie has input 0.}}

In this case, the first CGU$_\text{2N}$ in the inner chain is in working mode 4, while the second and third CGU$_\text{2N}$s are in working mode 1. In an inner chain, the photon evolution still obeys $\left|001\right\rangle\rightarrow\cos^{6N}\left(\pi/2N\right)|001\rangle$. As a result, Alice's output is 0 when $M\rightarrow\infty$. Regarding the probability of $D_0$ clicking, i.e., $P_{10D0}$, it is equal to $P_{01D0}$ and $P_{11D0}$. 

\begin{figure}[htbp]
	\centering\includegraphics[width=1.0\columnwidth]{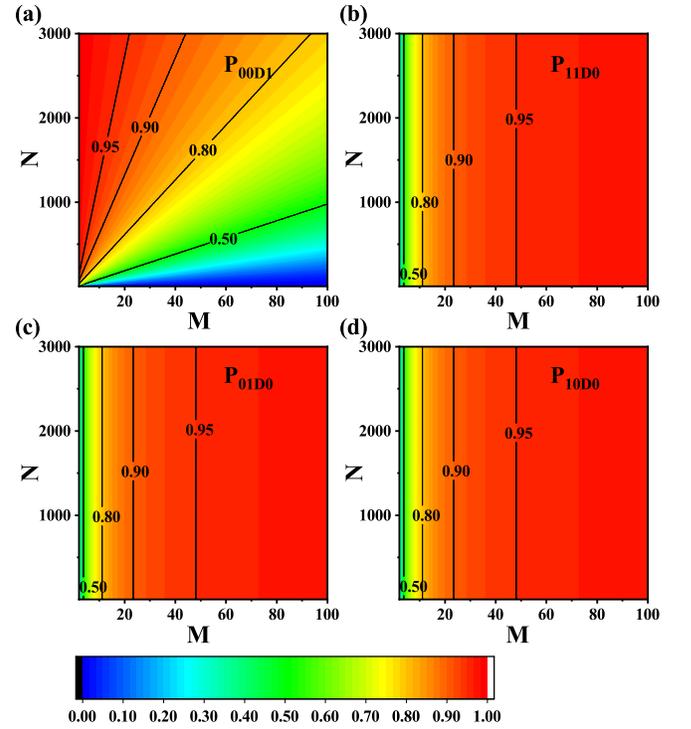}
	\caption
	{ \label{fig:fig5}
		Numerical simulation results for the counterfactual NOR gate. (a) $P_{00D1}$ represents the probability of $D_1$ clicking when both Bob and Charlie have input 0. (b) $P_{11D0}$ represents the probability of $D_0$ clicking when both Bob and Charlie have input 1. (c) $P_{01D0}$ represents the probability of $D_0$ clicking when Bob has input 0, while Charlie has input 1. (d) $P_{10D0}$ represents the probability of $D_0$ clicking, when Bob has input 1 and Charlie has input 0.}
\end{figure}

\begin{figure}[htbp]
	\centering\includegraphics[width=1.0\columnwidth]{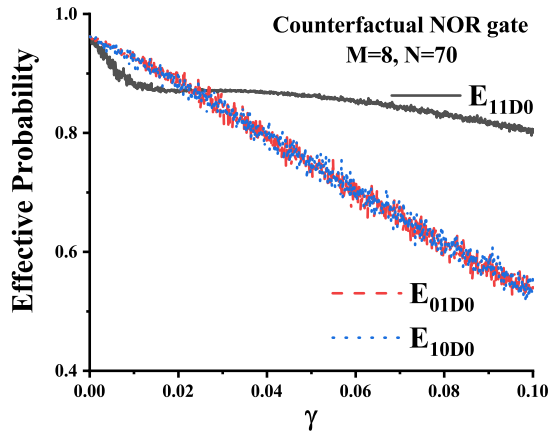}
	\caption
	{ \label{fig:fig6}
		Numerical simulation of the influence of channel noise on the counterfactual NOR gate. It is assumed that the transmission channels are blocked by some objects other than Bob and Charlie with probability $\gamma$. $E_{11D0}$ represents the effective probability of $D_0$ clicking, when both Bob and Charlie have input 1. $E_{01D0}$ represents the effective probability of $D_0$ clicking, when Bob has input 0, while Charlie has input 1. $E_{10D0}$ represents the effective probability of $D_0$ clicking when Bob has input 1, but Charlie has input 0. The curves of  $E_{01D0}$ and  $E_{10D0}$ are almost identical.}
\end{figure}

Consequently, the truth table for NOR gate can be implemented.

\subsection{Analysis of counterfactuals}

In case that both Bob and Charlie select to block their transmission channels, any photons entering the transmission channels must be detected by $SW$s. Since the single photon is used, we can see that when $D_1$ clicks, the photon never enters the transmission channels. In case that either Bob or Charlie selects to unblock his own transmission channel, we can consider each middle interferometer and its corresponding $D_3$ as a combined detector. We emphasize that due to the attenuators, the two arms of each middle interferometer are always balanced for cases 2, 3 and 4, even when $N$ is finite. In such cases, once a photon enters a middle interferometer, it is either absorbed by attenuators and $D_2$s or routed to $D_3$. The photon never remains in the outer chain. As a result, when $D_0$ clicks, it is an indication that the photon never enters the middle interferometers. We emphasize that the transmission channels are located inside the middle interferometers. Thus, the photon never enters the transmission channels. Consequently, the presented NOR gate is counterfactual.

In addition, here it's worth mentioning one difference between the design of the NOR gate and the SLAZ protocol. In the SLAZ protocol, the path evolution of a single photon inside Alice's optical device is interrupted by the status of the transmission channel (either blocked or unblocked). Such interruption occurs many times during the dynamic evolution of the photon, and all of those interruptions are of the same type (blocking or unblocking). However, for the counterfactual NOR gate, the situation is different. The statuses of the transmission channel during the dynamic evolution of the photon are no longer the same but rather mixed of blocked and unblocked. This is because the channel is now controlled by Bob and Charlie, and their actions are not related. Accordingly, the statuses of the transmission channel can form some specific time-order sequences. Based on that, we design Alice's optical device so that it can distinguish those specific sequences. Consequently, the current design of the NOR gate implies a different way to achieve counterfactual quantum control or communication.

\subsection{Numerical simulation}

To support our theory, we carry out numerical simulations, where no approximation is used. First, we consider the ideal case, except that the parameters $M$ and $N$ are finite, which means the number of $BS$ we used is limited. In such case, the probabilities of Alice getting correct result according to different Bob and Charlie's actions, i.e., $P_{00D1}$, $P_{11D0}$, $P_{01D0}$ and $P_{10D0}$, depend only on $M$ and $N$. The corresponding simulation results are given in Fig.\ref{fig:fig5}, which are in line with our expectations. As $M$ and $N$ increase, the probability of Alice getting the correct output increases and tends to 1. As for the cases that either Bob or Charlie has input 1, the numerical simulation results are the same. They are independent of $N$. Considering the case $M=30$ and $N=2500$, the numerical result is $P_{11D0}=P_{01D0}=P_{10D0}=0.921$, while the theoretical result is $0.918$. Clearly, the two results are in good agreement with each other. Same can be seen for Fig.\ref{fig:fig5}(a), the numerical result for $M=30$ and $N=2500$ is $0.918$, while the theoretical result is $0.911$. Therefore, our theoretical analyses are consistent with the numerical results.

In the above, we assume that all experimental equipment and conditions are perfect. However, we note that the SLAZ-based communication protocols are very sensitive to channel noise \cite{ours2,ouroe}, i.e., the transmission channels may be blocked by some other objects randomly, but neither by Bob or Charlie. Therefore, we run numerical simulations to see the effect of channel noise. In particular, we focus on the effective probability of the correct Alice's detector clicking, which is defined as $E_{jj'Dq}=P_{jj'Dq}/(P_{jj'D0}+P_{jj'D1}) $($j,j',q=0,1$). Here, $P_{jj'Dq}$ represents the probability of $D_q$ clicking when Bob has input $j$ and Charlie has input $j'$. We discard the cases where Alice does not get any output (neither $D_0$ or $D_1$ clicks).

 We plot Fig.6 for $E_{11D0}$ (black solid line), $E_{01D0}$ (red dashed line) and $E_{10D0}$ (blue dotted line) versus $\gamma$ with $M=8$ and $N=70$, where  $\gamma$ is defined as the probability of the transmission channels being blocked unexpectedly \cite{ours2,ouroe}. For each $\gamma$ value, we take multiple samples and calculate the average effective probabilities. 
 $E_{00D1}$ is not plotted since it is not affected by channel noise, whose numerical result is 91.5\%. As shown in the figure, when $\gamma$ is lower than 3\%, the effective probabilities $E_{11D0}$, $E_{01D0}$ and $E_{10D0}$ are all above 80\%.

\section{Counterfactual XOR gate}

In this section, we introduce the counterfactual XOR gate, and its corresponding truth table can be seen in Fig.\ref{fig:fig1}. 

\subsection{Scheme}

\begin{figure}[htbp]
	\centering\includegraphics[width=0.8\columnwidth]{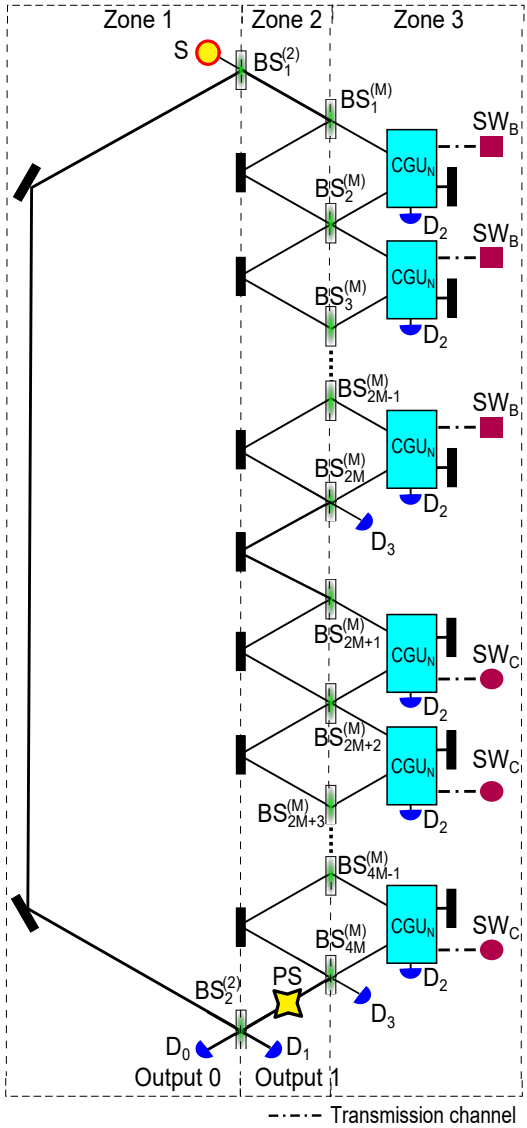}
	\caption
	{ \label{fig:fig7}
		The scheme of the counterfactual XOR gate. The details of CGU$_\text{N}$ can be found in Fig.\ref{fig:fig2}.}
\end{figure}

In Fig.\ref{fig:fig7}, we present the detailed scheme, which has the structure of triple chained interferometers. The biggest Mach-Zehnder interferometer has two $BS^{(2)}$s, and we call it the outer interferometer. On the right arm of the outer interferometer, there are chained interferometers containing $4M$ $BS^{(M)}$s, and we call them the middle chain. In the middle chain, there are $4M-2$ middle interferometers. We call the first $2M-1$ middle interferometers the upper half-chain (from $BS^{(M)}_1$ to $BS^{(M)}_{2M}$), while the last $2M-1$ middle interferometers the lower half-chain (from $BS^{(M)}_{2M+1}$ to $BS^{(M)}_{4M}$). The two half-chains have the same optical structure but are controlled by Bob and Charlie, respectively. In addition, in the right arm of each middle interferometer, there is a CGU$_{\text{N}}$.

For clarity, we divide Fig.\ref{fig:fig7} into three zones. The photon state $|100\rangle$ represents a photon in Zone 1 (the left side of $BS^{(2)}$s). The state $|010\rangle$ represents a photon in Zone 2 (the area between $BS^{(2)}$s and $BS^{(M)}$s), and the state $|001\rangle$ represents a photon in Zone 3 (the right side of $BS^{(M)}$s). 

\begin{figure}[htbp]
	\centering\includegraphics[width=1.0\columnwidth]{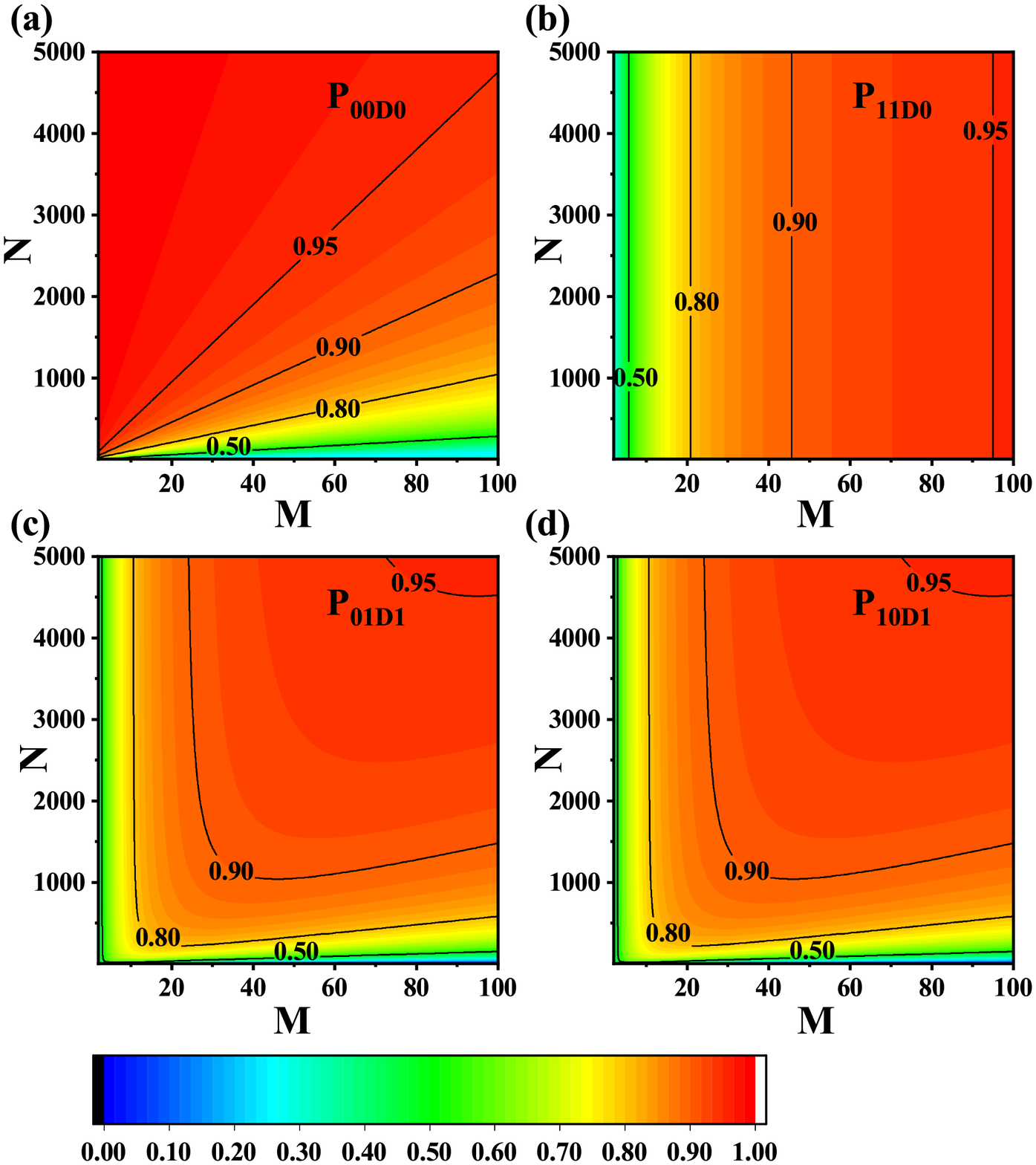}
	\caption
	{ \label{fig:fig8}
		Numerical simulation results for the counterfactual XOR gate. (a) $P_{00D0}$ represents the probability of $D_0$ clicking, when both Bob and Charlie have input 0. (b) $P_{11D0}$ represents the probability of $D_0$ clicking, when both Bob and Charlie have input 1. (c) $P_{01D1}$ represents the probability of $D_1$ clicking when Bob has input 0 but Charlie has input 1. (d) $P_{10D1}$ represents the probability of $D_1$ clicking when Bob has input 1 while Charlie has input 0.}
\end{figure}

\begin{figure}[htbp]
	\centering\includegraphics[width=1.0\columnwidth]{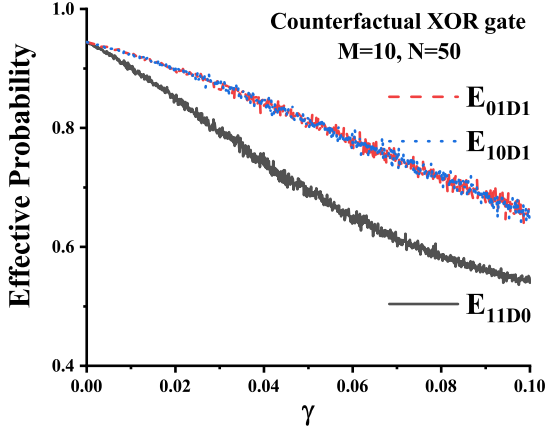}
	\caption
	{ \label{fig:fig9}
		Numerical simulation of channel noise influence on the counterfactual XOR gate. It is assumed that the transmission channels are blocked by some objects other than Bob and Charlie with the probability $\gamma$. $E_{11D0}$ represents the effective probability of $D_0$ clicking, when both Bob and Charlie have input 1. $E_{01D1}$ is the effective probability of $D_1$ clicking, when Bob has input 0 but Charlie has input 1. $E_{10D1}$ is the effective probability of $D_1$ clicking when Bob has input 1 while Charlie has input 0. The curves of  $E_{01D1}$ and  $E_{10D1}$ are almost identical.}
\end{figure}

Before getting into more detailed discussion, we first consider the evolution of a single photon in a half-chain.

Firstly, we assume that all CGU$_{\text{N}}$s are in the unblocking case, i.e., they are in working mode 2. In this case, any photons entering a CGU$_{\text{N}}$ must be routed to $D_2$ and absorbed. This leads the half-chain to be in working mode 1, and the function of the half-chain can be described as $\left|010\right\rangle\rightarrow\sqrt{{1-P}_1}|010\rangle$, where $P_1=1-\cos^{4M}\left(\pi/2M\right)\approx\pi^2/2M$ is the photon loss probability of the half-chain for  the unblocking case. 

Secondly, we consider the situation that all CGU$_{\text{N}}$s are in the blocking case, i.e., now they are in working mode 1. Under the condition that $N$ is sufficiently large, the photon is very likely to remain in the half-chain. Now, the half-chain is in working mode 3. Similar to the calculations of Eqs. \eqref{e3} and \eqref{e5}, here we can estimate the photon loss probability of the half-chain in the blocking case, which is $P_0=\sum_{m=1}^{2M}{{\sin}^2\left(\pi m/2M\right)}\left[{1-\cos}^{2N}\left(\pi/2N\right)\right]\approx\pi^2M/4N$. The function of the half-chain can be roughly written as  $\left|010\right\rangle\rightarrow-\sqrt{{1-P}_0}\left|010\right\rangle$. Comparing to the unblocking case, a $\pi$ phase shift occurs.

In summary, when a single photon in state $\left|010\right\rangle$ is sent into a half-chain, the final state is  ${(-1)}^{j+1}\sqrt{1-P_j}\left|010\right\rangle$ with $j=0,1$ for the blocking and unblocking cases, respectively. 

Now, we explain how the counterfactual XOR gate works. Initially, a single photon in state $\left|100\right\rangle$ is sent into the counterfactual XOR gate. After passing through $BS^{(2)}_1$, the photon state becomes $(|100\rangle+|010\rangle)/\sqrt{2}$. The photon in state $\left|100\right\rangle$ remains in the left arm of the outer interferometer, while the photon in state $\left|010\right\rangle$ enters the middle chain. Considering that the middle chain includes two half-chains and a $PS$, the photon evolution in the middle chain can be described as $\left|010\right\rangle\rightarrow{-(-1)}^{j+j'}\sqrt{1-P_j}\sqrt{1-P_{j'}}\left|010\right\rangle$, where the ``$-$'' comes from $PS$, and $j,j'=0,1$ represents Bob and Charlie's inputs, respectively. Then, after passing through $BS^{\left(2\right)}_2$, Alice's photon state becomes
\begin{equation}
\begin{split}
\label{e8}
&\frac{1}{2}\left[1{+\left(-1\right)}^{j+j'}\sqrt{1-P_j}\sqrt{1-P_{j'}}\right]\left|100\right\rangle \\
&+\frac{1}{2}\left[1{-\left(-1\right)}^{j+j'}\sqrt{1-P_j}\sqrt{1-P_{j'}}\right]\left|010\right\rangle \\
\end{split}
\end{equation}
In particular, when $N\gg M\rightarrow\infty$, the photon state is
\begin{equation}
\begin{split}
\label{e9}
\frac{1}{2}\left[1{+\left(-1\right)}^{j+j'}\right]\left|100\right\rangle+\frac{1}{2}\left[1{-\left(-1\right)}^{j+j'}\right]\left|010\right\rangle
\end{split}
\end{equation}
Eq.\eqref{e9} indicates that if Bob and Charlie have the same inputs, $D_0$ clicks (output 0), while if they have different inputs, $D_1$ clicks (output 1). Hence, the XOR gate is achieved. When $M$ and $N$ are finite, we have the following results.

{\bf{Case 1: Both Bob and Charlie have input 0.}} 

Here, $D_0$ clicks with the probability $P_{00D0}=\left|2-P_0\right|^2/4\approx1-\pi^2M/4N$. 

{\bf{Case 2: Both Bob and Charlie have input 1.}}

In this case, $D_0$ clicks with the probability $P_{11D0}=\left|2-P_1\right|^2/4\approx1-\pi^2/2M$.

{\bf{Case 3\&4: Bob and Charlie have different inputs.}}

For such cases, $D_1$ clicks with the probability $P_{01D1}=P_{10D1}=\left|1+\sqrt{1-P_0}\sqrt{1-P_1}\right|^2/4\approx1-\pi^2M/8N-\pi^2/4M$.

\subsection{Analysis of counterfactuals}

	Here the transmission channels are in the right arms of the CGU$_{\text{N}}$s. Considering a CGU$_{\text{N}}$, we assume that a photon is sent into it initially. If the transmission channel is blocked, the photon entering the transmission channel must be absorbed by $SW$s so it cannot remain in the CGU$_{\text{N}}$. If the transmission channel is unblocked, however, the photon must be routed to $D_2$ and absorbed. The photon cannot remain in Alice's device. Consequently, regardless of what Bob and Charlie's actions are, when either $D_0$ or $D_1$ clicks, the photon never enters the transmission channels. This leads to the conclusion that the presented XOR gate is counterfactual.

\subsection{Numerical simulation}

Again in order to support our theoretical analyses, we show  numerical simulation results in Fig.\ref{fig:fig8} without any approximation, where $P_{00D0}$, $P_{11D0}$, $P_{01D1}$ and $P_{10D1}$ are plotted. As shown in the figure, when $M$ and $N$ increase, Alice's probability of getting correct output increases and tends to 1. To compare our numerical results with the theoretical analyses, we consider the case $M=100$ and $N=3000$. For Fig.\ref{fig:fig8}(a),  the numerical result is $P_{00D0}=0.923$, while it is $0.918$ theoretically.  For Fig.\ref{fig:fig8}(b), numerical result gives $P_{11D0}=0.952$, while it is $0.951$ theoretically.   For Fig.\ref{fig:fig8}(c) and (d), the numerical result is $P_{01D1}=P_{10D1}=0.937$, while the theoretical result gives $0.934$. All these results are in good agreement with each other.

Furthermore, we plot Fig.9 to show the influence of channel noise. We plot $E_{11D0}$ (black solid line), $E_{01D1}$ (red dashed line) and $E_{10D1}$ (blue dotted line) versus $\gamma$ with $M=10$ and $N=50$. The numerical simulation results show that when $\gamma$ is around 3\%, $E_{11D0}$ is around 80\%, while $E_{01D1}$ and $E_{10D1}$ are around 87\%. As for $E_{00D0}$, it is not affected by channel noise and its numerical simulation result is 94.4\%.

\section{Counterfactual multi-party entanglement}

In all the above counterfactual logic gates, the transmission channels are controlled using classical objects. One would ask what happens if these classical objects are replaced by quantum objects, which are in superposition states of blocking and unblocking the transmission channels? In this section, we show that this was achieved before, but for just one party and that is Bob, and the results are not difficult to obtain according to \cite{tele0,ours4,ourpra}. We show that three remote quantum objects held by Bob, Charlie and David can be counterfactually entangled with the assistance of Alice's photon, thereby achieving GHZ state or W state.

\begin{figure}[htbp]
	\centering\includegraphics[width=1.0\columnwidth]{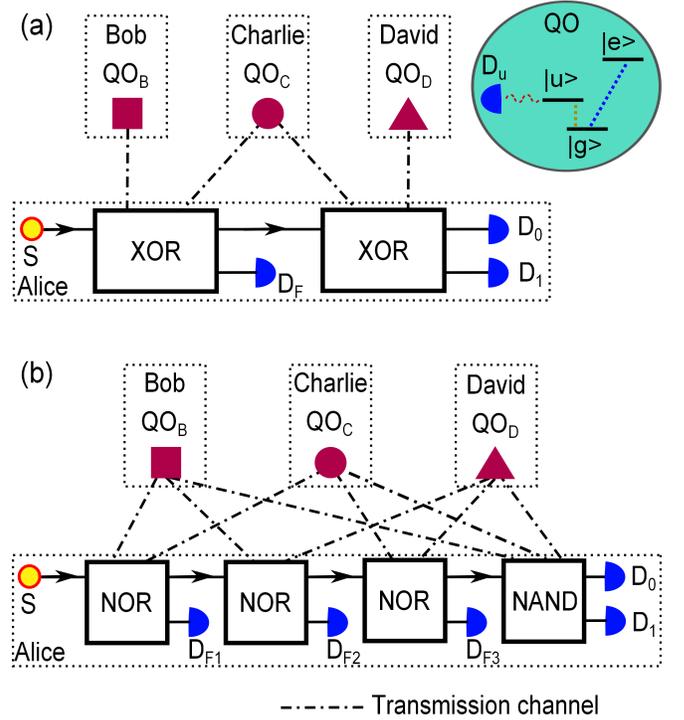}
	\caption
	{ \label{fig:fig10}
		The preparation of (a) GHZ state and (b) W state utilizing counterfactual logic gates. $QO$ stands for quantum object, which can be a V-type three-level atom with ground state $|g\rangle$ and two excited states $|u\rangle$ and $|e\rangle$. Only the transition between $|g\rangle$ and  $|u\rangle$ is coupled by Alice's photon. We assume that if the atom jumps to state $|u\rangle$, it must alert the detector $D_u$. Utilizing counterfactual logic gates (NOR, XOR and NAND gates), Bob, Charlie and David's $QO$s can be entangled counterfactually.}
\end{figure}

Here, for the simplicity, we utilize a V-type three-level atom as the quantum object, which is shown in Fig.\ref{fig:fig10}, where $|g\rangle$ is the ground state, $|e\rangle$ and $|u\rangle$ are two excited states. Only the transition between the atomic states $|g\rangle$ and $|u\rangle$ is coupled by Alice's photon. Consider the case where the atom is placed in the transmission channel and Alice's photon also appears in the channel. If the atom is in state $|g\rangle$, it absorbs Alice's photon and jumps to $|u\rangle$. We assume that this must cause $D_u$ to click. As a result, the transmission channel is blocked. The opposite scenario is that when the atom is in state $|e\rangle$, it has no influence on Alice's photon. Then, the photon just passes through the atom and is reflected by a mirror (not shown in the figure), which corresponds to the unblocking case. Obviously, when the atom is prepared in a superposition state of $| g\rangle$ and $| e\rangle $, the transmission channel is in a superposition state of being blocked and unblocked.  

Before getting into some discussions, we need to emphasize two issues. First, the V-type three-level atom is a good model for illustration, while generic experiment need better and realistic candidates such as  a Rydberg atom \cite{tele0,Atom1,Atom2} or a single-side cavity with a three-level atom \cite{ours4,qn27,qn28,inputout}. Second, in the presented counterfactual logic gates, Bob (or Charlie) must manipulate Alice's photon many times. The key point here is that Bob (or Charlie) must use the same quantum object to manipulate Alice's photon all the time. That requires the quantum object to be movable from one interferometer to another. Hence, better alternative to go around this issue is to consider using  Michelson interferometer, which is not difficult to achieve according to previous results \cite{ours2,ours4,ourpra}.

Next, we discuss the counterfactual NOR, XOR and NAND gates one by one. We assume that initially, Bob, Charlie and David each have an atom prepared in an arbitrary state $C^{(g)}_s| g\rangle +C^{(e)}_s| e\rangle $, where $C^{(g)}_s$ and $C^{(e)}_s$ are probability amplitudes satisfying $| C^{(g)}_s|^2+|C^{(e)}_s|^2=1$. Here the subscript $s=B$ indicates that the atom belongs to Bob, while $s=C,D$ indicates Charlie and David, respectively. As for Alice, we emphasize that for each type of counterfactual logic gate, there is only one photon entrance. Therefore, we just set initial Alice's photon state to be $|i_A\rangle$. As for the state of the photon after passing through the logic gate, we assume that $|0(1)_A\rangle$ represents Alice's photon appearing at output 0(1) (see Figs.\ref{fig:fig3}, \ref{fig:fig4} and \ref{fig:fig7}).

First, we consider the NOR gate. Initially, the whole atom-photon state is
\begin{equation}
\begin{split}
\label{f01}
&|i_A\rangle (C^{(g)}_B | g_B \rangle +C^{(e)}_B | e_B \rangle )(C^{(g)}_C | g_C \rangle +C^{(e)}_C | e_C \rangle) \\
&=C^{(g)}_B C^{(g)}_{C}|i_A\rangle  | g_B \rangle  | g_C \rangle + C^{(e)}_BC^{(e)}_C|i_A\rangle  | e_B \rangle  | e_C \rangle \\
&+C^{(g)}_B C^{(e)}_C|i_A\rangle  | g_B \rangle  | e_C \rangle + C^{(e)}_BC^{(g)}_C|i_A\rangle  | e_B \rangle  | g_C \rangle \\
\end{split}
\end{equation} 
We emphasize that during the information processing, Bob and Charlie's atomic states do not change unless one of the atom absorbs Alice's photon, which leads to $|g\rangle \to |u\rangle$. As a result, the four terms in the second and third lines of Eq. \eqref{f01} are always orthogonal to each other no matter what the evolution of Alice's photon is. They indicate four subsystems \cite{ours4} and the photon evolutions in these subsystems are independent. Moreover, we notice that $|e\rangle$ means unblocking while $|g\rangle$ means blocking. The evolution of Alice's photon obeys the rules presented in Section V. After the NOR gate and in the ideal case ($N \gg M \to \infty$), Eq.\eqref{f01} becomes 
\begin{equation}
\begin{split}
\label{e10}
\overset{NOR}{\to}& C^{(g)}_B C^{(g)}_C|1_A\rangle  | g_B \rangle  | g_C \rangle + C^{(e)}_BC^{(e)}_C|0_A\rangle  | e_B \rangle  | e_C \rangle \\
&+C^{(g)}_B C^{(e)}_C|0_A\rangle  | g_B \rangle  | e_C \rangle + C^{(e)}_BC^{(g)}_{C}|0_A\rangle  | e_B \rangle  | g_C \rangle \\
\end{split}
\end{equation}

Similar results can also be found for the counterfactual XOR and NAND gates. For the counterfactual XOR gate, we have
\begin{equation}
\begin{split}
\label{e11}
&|i_A\rangle (C^{(g)}_B | g_B \rangle +C^{(e)}_B | e_B \rangle )(C^{(g)}_C | g_C \rangle +C^{(e)}_C | e_C \rangle) \\
&\overset{XOR}{\to} C^{(g)}_B C^{(g)}_{C}|0_A\rangle  | g_B \rangle  | g_C \rangle + C^{(e)}_BC^{(e)}_C|0_A\rangle  | e_B \rangle  | e_C \rangle \\
&+C^{(g)}_B C^{(e)}_C|1_A\rangle  | g_B \rangle  | e_C \rangle + C^{(e)}_BC^{(g)}_C|1_A\rangle  | e_B \rangle  | g_C \rangle \\
\end{split}
\end{equation}

Regarding the M-type counterfactual NAND gate that is controlled by Bob, Charlie and David, we have
\begin{equation}
\begin{split}
\label{e12}
  & \left| i_A \right\rangle \left( C^{(g)}_B\left| g_B \right\rangle +C^{(e)}_B\left| e_B \right\rangle  \right) \\ 
& \times \left( C^{(g)}_C \left| g_C \right\rangle +C^{(e)}_C\left| e_C \right\rangle  \right)\left( C^{(g)}_D\left| g_D \right\rangle +C^{(e)}_D\left| e_D \right\rangle  \right) \\ 
& \overset{NAND}{\mathop{\to }}\,C^{(e)}_B C^{(e)}_CC^{(e)}_D\left| 0_A \right\rangle \left| e_B \right\rangle \left| e_C \right\rangle \left| e_D \right\rangle  \\ 
 & - C^{(e)}_B C^{(e)}_C C^{(e)}_D\left| 1_A \right\rangle \left| e_B \right\rangle \left| e_C \right\rangle \left| e_D \right\rangle \\
 &+\left| 1_A \right\rangle \left( C^{(g)}_B \left| g_B \right\rangle + C^{(e)}_B\left| e_B \right\rangle  \right) \\ 
& \times \left( C^{(g)}_C\left| g_C \right\rangle +C^{(e)}_C\left| e_C \right\rangle  \right)\left( C^{(g)}_D \left| g_D \right\rangle +C^{(e)}_D\left| e_D \right\rangle  \right) \\ 
\end{split}
\end{equation}
In other words, only in the case when Bob, Charlie and David have input 1, the output for Alice's photon is $\left| 0_A \right\rangle $. For all other initial states related to the M-type counterfactual NAND gate, the final state of Alice's photon is always $\left| 1_A \right\rangle $.

In the following, based on Eqs.\eqref{e10}\eqref{e11}\eqref{e12},  we show how to use the counterfactual logic gates to counterfactually entangle three remote quantum objects held by Bob, Charlie and David, and thereby achieving three-qubit GHZ state $(\left| g_B \right\rangle \left| g_C \right\rangle \left| g_D \right\rangle +\left| e_B \right\rangle \left| e_C \right\rangle \left| e_D \right\rangle )/\sqrt{2}$ and W state $(\left| g_B \right\rangle \left| g_C \right\rangle \left| e_D \right\rangle +\left| g_B \right\rangle \left| e_C \right\rangle \left| g_D \right\rangle +\left| e_B \right\rangle \left| g_C \right\rangle \left| g_D \right\rangle )/\sqrt{3}$. We assume that all optical and atomic systems are perfect.

\subsection{ Preparation of three-qubit GHZ state}

In Fig.\ref{fig:fig10}(a), we present a scheme on how to generate a GHZ state counterfactually. Initially, Bob, Charlie and David prepare their atoms in the same superposition state $\left( \left| e \right\rangle +\left| g \right\rangle  \right)/\sqrt{2}$. The state of the whole system is 
\begin{equation}
\begin{split}
\label{e13}
\left| i_A \right\rangle \frac{\left( \left| g_B \right\rangle +\left| e_B \right\rangle  \right)}{\sqrt{2}}\frac{\left( \left| g_C \right\rangle +\left| e_C \right\rangle  \right)}{\sqrt{2}}\frac{\left( \left| g_D \right\rangle +\left| e_D \right\rangle  \right)}{\sqrt{2}}
\end{split}
\end{equation}

At Alice's end, she sends her single photon into the first counterfactual XOR gate, which is controlled by Bob and Charlie. According to Eq.\eqref{e11}, the whole system state evolves to
\begin{equation}
\begin{split}
\label{e14}
&\frac{1}{2\sqrt{2}}\left| 0_A \right\rangle \left( \left| g_B \right\rangle \left| g_C \right\rangle +\left| e_B \right\rangle \left| e_C \right\rangle  \right)\left( \left| g_D \right\rangle +\left| e_D \right\rangle  \right) \\
&+\frac{1}{2\sqrt{2}}\left| 1_A \right\rangle \left( \left| g_B \right\rangle \left| e_C \right\rangle +\left| e_B \right\rangle \left| g_C \right\rangle  \right)\left( \left| g_D \right\rangle +\left| e_D \right\rangle  \right) \\
\end{split}
\end{equation}
The photon in the path of output 1 ($\left| 1_A \right\rangle $) is measured by $D_F$. If $D_F$ clicks, the entanglement process fails. If $D_F$ does not click, the remaining system state is $ \left| 0_A \right\rangle ( \left| g_B \right\rangle \left| g_C \right\rangle +\left| e_B \right\rangle \left| e_C \right\rangle )( \left| g_D \right\rangle +\left| e_D \right\rangle) / {2\sqrt{2}}$ without renormalization. The surviving photon enters the second counterfactual XOR gate, which is controlled by Charlie and David. After the gate, the whole system state becomes
\begin{equation}
\begin{split}
\label{e15}
&\frac{1}{2\sqrt{2}}\left| 0_A \right\rangle \left( \left| g_B \right\rangle \left| g_C \right\rangle \left| g_D \right\rangle +\left| e_B \right\rangle \left| e_C \right\rangle \left| e_D \right\rangle  \right) \\
&\text{+}\frac{1}{2\sqrt{2}}\left| 1_A \right\rangle \left( \left| g_B \right\rangle \left| g_C \right\rangle \left| e_D \right\rangle +\left| e_B \right\rangle \left| e_C \right\rangle \left| g_D \right\rangle  \right) \\
\end{split}
\end{equation}
The photon in state $\left| {{\text{0}}_{A}} \right\rangle $ is measured by $D_0$, while the photon in state $\left| 1_A \right\rangle $ is measured by $D_1$.  Apparently, when $D_0$ clicks, the state of the three quantum objects must be $\left( \left| g_B \right\rangle \left| g_C \right\rangle \left| g_D \right\rangle +\left| e_B \right\rangle \left| e_C \right\rangle \left| e_D \right\rangle  \right)/\sqrt{2}$, while the probability of successfully generating the GHZ state counterfactually is 25\%.

\subsection{ Preparation of W state}

Fig.\ref{fig:fig10}(b) shows the scheme that can generate W state counterfactually. Initially, Bob, Charlie and David still prepare their atoms in the same superposition state $\left( \left| e \right\rangle +\left| g \right\rangle  \right)/\sqrt{2}$. At Alice's end, she sends her single photon into the first counterfactual NOR gate, which is controlled by Bob and Charlie. According to Eq.\eqref{e10}, the whole system state evolves to
\begin{equation}
\begin{split}
\label{e16}
& \frac{1}{2\sqrt{2}}\left| 1_A \right\rangle \left| g_B \right\rangle \left| g_C \right\rangle \left( \left| g_D \right\rangle +\left| e_D \right\rangle  \right) +\frac{1}{2\sqrt{2}}\left| 0_A \right\rangle\\ 
&  \times \left( \left| g_B \right\rangle \left| e_C \right\rangle +\left| e_B \right\rangle \left| g_C \right\rangle +\left| e_B \right\rangle \left| e_C \right\rangle  \right)\left( \left| g_D \right\rangle +\left| e_D \right\rangle  \right) \\ 
\end{split}
\end{equation}
Then, the photon in state $\left| 1_A \right\rangle $ is measured by $D_{F1}$, and if $D_{F1}$ clicks, it is an indication of entanglement failure. If $D_{F1}$ does not click, the surviving photon is routed to the second counterfactual NOR gate, which is controlled by Bob and David. After the second counterfactual NOR gate, the system state becomes
\begin{equation}
\begin{split}
\label{e17}
& \frac{1}{2\sqrt{2}}\left| 1_A \right\rangle \left| g_B \right\rangle \left| e_C \right\rangle \left| g_D \right\rangle  \\ 
& +\frac{1}{2\sqrt{2}}\left| 0_A \right\rangle \left( \left| g_B \right\rangle \left| e_C \right\rangle \left| e_D \right\rangle +\left| e_B \right\rangle \left| g_C \right\rangle \left| g_D \right\rangle \right. \\
& \left. +\left| e_B \right\rangle \left| g_C \right\rangle \left| e_D \right\rangle +\left| e_B \right\rangle \left| e_C \right\rangle \left| g_D \right\rangle +\left| e_B \right\rangle \left| e_C \right\rangle \left| e_D \right\rangle  \right) \\ 
\end{split}
\end{equation} 
Now, the photon in state $|1_A\rangle$ is measured by $D_{F2}$. If $D_{F2}$ does not click, then the surviving photon is sent to the third counterfactual NOR gate, which is controlled by Charlie and David. After the third counterfactual NOR gate, the system state becomes
\begin{equation}
\begin{split}
\label{e18}
& \frac{1}{2\sqrt{2}}\left| 1_A \right\rangle \left| e_B \right\rangle \left| g_C \right\rangle \left| g_D \right\rangle  \\ 
& +\frac{1}{2\sqrt{2}}\left| 0_A \right\rangle \left( \left| g_B \right\rangle \left| e_C \right\rangle \left| e_D \right\rangle +\left| e_B \right\rangle \left| g_C \right\rangle \left| e_D \right\rangle \right. \\
&\left. +\left| e_B \right\rangle \left| e_C \right\rangle \left| g_D \right\rangle +\left| e_B \right\rangle \left| e_C \right\rangle \left| e_D \right\rangle  \right) \\ 
\end{split}
\end{equation}  
This time, $D_{F3}$ eliminates the photon in state $\left| 1_A \right\rangle $. The photon in state $\left| 0_A \right\rangle $ is sent to a M-type counterfactual NAND gate, which is controlled by Bob, Charlie and David. After the NAND gate, we have
\begin{equation}
\begin{split}
\label{e19}
&\frac{1}{2\sqrt{2}}\left| 1_A \right\rangle \left( \left| g_B \right\rangle \left| e_C \right\rangle \left| e_D \right\rangle +\left| e_B \right\rangle \left| g_C \right\rangle \left| e_D \right\rangle +\left| e_B \right\rangle \left| e_C \right\rangle \left| g_D \right\rangle  \right) \\
&+\frac{1}{2\sqrt{2}}\left| 0_A \right\rangle \left| e_B \right\rangle \left| e_C \right\rangle \left| e_D \right\rangle \\
\end{split}
\end{equation} 
It is not difficult to see, when $D_1$ clicks, the W state is almost achieved. Bob, Charlie and David only need to flip their bits ($\left| e \right\rangle \leftrightarrow \left| g \right\rangle $) to get the state $( \left| e_B \right\rangle \left| g_C \right\rangle \left| g_D \right\rangle +\left| g_B \right\rangle \left| e_C \right\rangle \left| g_D \right\rangle +\left| g_B \right\rangle \left| g_C \right\rangle \left| e_D \right\rangle)/\sqrt{3}$. The probability that W state is generated successfully and counterfactually is 37.5\%.

\section{Discussion about experiment}

After discussing and showing how to achieve NAND, NOR and XOR logic gates counterfactually. One question of interest is how possible to carry out an actual experimental implementation bearing in mind all of the required interferometric structures? SLAZ, which as we pointed out can be used as logical NOT operation, was demonstrated in a beautiful experiment by Jian-Wei Pan's group in 2017 \cite{exp01}.  Another protocol is the quantum counterfactual communication without a weak trace \cite{2b}, again was implemented using a programmable nanophotonic processor, which is based on reconfigurable silicon-on-insulator waveguides that operate at Telecom wavelengths \cite{3b}. These are two examples just to illustrate the fact that despite having a number of interferometric structures, but we can still have proof of principle experiments for these counterfactual quantum protocols. Needless to say that there is growing interest to have most if not all the quantum protocols implemented on chips.   One of the growing technology is the photonic integrated circuit architecture for a quantum programmable gate array \cite{4b}, this can be a good candidate to implement the aforementioned counterfactual gates.  Clearly, what we have proposed is not far reach from being done experimentally within the scope of present technology.

\section{Conclusion}

In this work, we show that by independently controlling the blocking and unblocking of the transmission channels, two remote parties (Bob and Charlie) can completely determine the state of a real photon that is confined to a third party's (Alice) station without any need for pre-arranged entanglement nor any real physical particles traveling among these three parties. Based on that, we show how to implement basic logic gates including NAND, NOR and XOR gates counterfactually. It is Bob and Charlie's actions that determine the gate inputs, while Alice's photon state determines the gate outputs. The nature of the chosen counterfactual logic gates depends only on the specific design of Alice's device. In addition, we show that utilizing the proposed counterfactual logic gates, we can counterfactually entangle three remote quantum objects and hence realize GHZ state and W state.

\begin{acknowledgments}
	This work is supported by National Natural Science Foundation of China (NSFC) (No. 11704241); This work is also supported by a grant from the King Abdulaziz City for Science and Technology (KACST).
	
\end{acknowledgments}


\end{document}